\documentclass[12pt,preprint]{aastex}
\usepackage{pslatex}

\begin{document}
\title{Radio-wave propagation through a medium containing 
electron-density fluctuations described by an
anisotropic  Goldreich-Sridhar spectrum}

\author{B. D. G. Chandran}
\affil{Department of Physics \& Astronomy, University of Iowa, IA;
benjamin-chandran@uiowa.edu}
\author{D. C. Backer}
\affil{ Astronomy Department \& Radio Astronomy Laboratory,
University of California, Berkeley, CA; dbacker@astro.berkeley.edu }

\begin{abstract}

We study the propagation of radio waves through a medium possessing
density fluctuations that are elongated along the ambient magnetic
field and described by an anisotropic Goldreich-Sridhar power
spectrum.  We derive general formulas for the wave phase structure
function $D_\phi$, visibility, angular broadening, diffraction-pattern
length scales, and scintillation time scale for arbitrary
distributions of turbulence along the line of sight, and specialize
these formulas to idealized cases.  In general, $D_\phi \propto
(\delta r) ^{5/3}$ when the baseline $\delta r$ is in the inertial
range of the turbulent density spectrum, and $D_\phi \propto (\delta
r) ^2$ when $\delta r$ is in the dissipation range, just as for an
isotropic Kolmogorov spectrum of fluctuations.  When the density
structures that dominate the scattering have an axial ratio $R \gg 1$
(typically $R \sim 10^3$), the axial ratio of the broadened image of a
point source in the standard Markov approximation is at most $\sim
R^{1/2}$, and this maximum value is obtained in the unrealistic case
that the scattering medium is confined to a thin screen in which the
magnetic field has a single direction. If the projection of the magnetic
field within the screen onto the plane of the sky rotates through an angle
$\Delta \psi$ along the line of sight from one side of the screen to
the other, and if $R^{-1/2} \ll \Delta \psi \ll 1$, then the axial
ratio of the resulting broadened image of a point source is
$2(8/3)^{3/5}/\Delta \psi \simeq 3.6/\Delta \psi$. The error in this
formula increases with $\Delta \psi$, but reaches only $\sim15$\% when
$\Delta \psi = \pi$.  This indicates that a moderate amount of
variation in the direction of the magnetic field along the line of
sight can dramatically decrease the anisotropy of a broadened
image. When $R\gg 1$, the observed anisotropy will in general be
determined by the degree of variation of the field direction along the
sight line and not by the degree of density anisotropy. Although this
makes it difficult to determine observationally the degree of
anisotropy in interstellar density fluctuations, observed anisotropies
in broadened images provide general support for anisotropic models of
interstellar turbulence.  Regions in which the angle $\gamma$ between
the magnetic field and line of sight is small cause enhanced
scattering due to the increased coherence of density structures along
the line of sight. In the exceedingly rare and probably unrealized
case that scattering is dominated by regions in which $\gamma \lesssim
(\delta r/l)^{1/3}$, where $l$ is the outer scale (stirring scale) of
the turbulence, $D_{\phi} \propto (\delta r)^{4/3}$ for $\delta r$ in
the inertial range.  In a companion paper (Backer \& Chandran)
we discuss the semi-annual modulation in the scintillation
time of a nearby pulsar for which the field-direction variation along
the line of sight is expected to be moderately small.

\end{abstract}

\section{Introduction}

Scattering of radio waves from point sources by electron-density
fluctuations in the interstellar medium (ISM) gives rise to a number
of effects, including intensity scintillation and
angular broadening (Rickett 1990). These phenomena provide a useful
diagnostic of density fluctuations in the ISM on scales $\sim 10^8 -
10^{10}$ cm (diffractive scintillation) and
$10^{13}- 10^{15}$ cm (refractive scintillation) (Armstrong et~al.\
1995). In a majority of cases, broadened images are anisotropic, with
axial ratios (long dimension of image divided by short dimension)
between 1.1 and 1.8 for sight lines through strongly scattering
regions in the Galactic disk (Mutel \& Lestrade 1990, Wilkinson
et~al.\ 1994, Molnar et~al.\ 1995, Spangler \& Cordes 1998, Trotter et
al. 1998) and as large as 3:1 for OH masers near the
Galactic center (van Langevelde et~al.\ 1992, Frail et~al.\ 1994).
[Axial ratios $> 5:1$ have been observed in the solar wind (Narayan et~al.\ 1990,
Armstrong et~al.\ 1990).]
This paper explores the relation between anisotropic scattering and
recent theories of anisotropic magnetohydrodynamic
(MHD) turbulence.

Early theories of MHD turbulence assumed isotropy. Iroshnikov (1963)
and Kraichnan (1965) independently derived a $k^{-3/2}$ power spectrum
for the velocity and magnetic field in isotropic incompressible
turbulence. Over the last decade, however, a number of authors have
investigated theories in which small-scale fluctuations are elongated
along the local direction of the magnetic field, ${\bf B}_0$ (e.g.,
Montgomery \& Turner 1981, Shebalin et~al.\ 1983, Higdon 1984, Higdon
1986, Oughton et~al.\ 1994, Sridhar \& Goldreich 1994, Goldreich \&
Sridhar 1995, Montgomery \& Matthaeus 1995, Ghosh \& Goldstein 1997,
Goldreich \& Sridhar 1997, Matthaeus et~al.\ 1998, Spangler 1999,
Bhattacharjee \& Ng 2000, Cho \& Vishniac 2000, Maron \& Goldreich
2001, Lithwick \& Goldreich 2001).  Using phenomenological arguments
and a statistical turbulence theory, Goldreich \& Sridhar (1995)
(hereafter GS) derived a form for the velocity and magnetic power
spectrum in incompressible MHD turbulence that approximately
corresponds to the anisotropic spectra found in direct numerical
simulations (Cho \& Vishniac 2000, Maron \& Goldreich 2001).  Lithwick
\& Goldreich (2001) extended the GS theory to compressible MHD
turbulence. They found that the spectra of shear-Alfv\'en modes, slow
modes, and entropy modes all have the same anisotropic form as the
shear-Alfv\'en and pseudo-Alfv\'en modes in incompressible turbulence.
The density fluctuations in the compressible turbulence are dominated
by the slow modes and entropy modes.  Lithwick \& Goldreich (2001)
considered the effects of damping by neutrals, radiative cooling,
electron heat conduction, and ion diffusion, and found that the
density spectrum can extend to the small scales responsible for
diffractive scintillation provided that the neutral fraction is very
small, and that either $\beta$ (the ratio of thermal to magnetic
pressure) is not much larger than 1 or the outer scale $l$ of the
turbulence is fairly small. For $\beta \sim 1$, the density spectrum
is cut off at wavelengths along the magnetic field comparable to the
proton mean free path. Because the wavelength of the density
fluctuations across the magnetic field is much smaller than the
parallel wavelength, the density spectrum extends to an inner
scale $d$ significantly smaller than the proton mean free path, of order
(Lithwick \& Goldreich 2001)
\begin{equation}
d \simeq 2 \times 10^9 \left(\frac{\mbox{ pc}}{l}\right)^{1/2} \left(
\frac{\beta}{n/\mbox{ cm}^{-3}}\right)^{3/2} \mbox{ cm}.
\label{eq:densitycutoff} 
\end{equation} 
The axial ratio (long dimension divided by short dimension) of
fluctuations of perpendicular scale $d$ in the GS spectrum is $\sim
(l/d)^{1/3}$, which is $\sim 10^3$ for $l= 1 $ pc, $\beta =1$, and $n=
1 \mbox{ cm}^{-3}$. 

In this paper, we take the interstellar density
fluctuations to have a GS spectrum 
and calculate the consequences for radio wave propagation
in the ISM. In section
\ref{sec:background} we review the relations between the wave phase
structure function, visibility, angular broadening,
diffraction-pattern length scales, and scintillation time scale. In
section \ref{sec:calc} we derive formulas for these quantities
 for arbitrary distributions of
turbulence along the line of sight, and specialize these formulas to
idealized observational scenarios. In section \ref{sec:conc} we give a
detailed summary of our main results.

\section{Background: wave phase structure function,
visibility,  angular  broadening, diffraction-pattern length
scales, and scintillation time scale}
\label{sec:background} 

In this section, some general results on scintillation and angular
broadening are reviewed. For an overview of the subject, the reader is
referred to Rickett (1990). A systematic derivation of the
interferometric visibility and intensity correlation function for a
plane wave propagating through a stationary medium given a stationary
observer was given by Lee \& Jokipii (1975a,b). 
Their derivation is
extended to the case of a moving point source and moving observer in
appendix~\ref{ap:deriv}.

The visibility, $\langle E(\vec{r},t) E^\ast(\vec{r} + \vec{\delta
r},t+\delta t) \rangle$, is the correlation between the electric field
observed at position $\vec{r}$ in the earth's reference frame at time
$t$ and the electric field at position $\vec{r} + \vec{\delta r}$ in
the earth's reference at time $t + \delta t$.  In the Markov
approximation (Lee \& Jokipii 1975a), which requires that as a wave
propagates through one correlation length of an electron density fluctuation the
change to the wave field induced by the density fluctuation is small,
the visibility is given by (Lotova \& Chashei 1981, Cordes \& Rickett
1998; see appendix~\ref{ap:deriv} for a derivation),
\begin{equation} 
\langle E(\vec{r}, t) E^\ast (\vec{r} + \vec{\delta r}, t+ \delta t) \rangle = 
\exp[-D_{\phi }(\vec{\delta r},\delta t)/2],
\label{eq:vis} 
\end{equation} 
where
\begin{equation}
D_\phi(\vec{\delta r},\delta t)=
4\pi r_{\rm e}^2\lambda^2\int_0^Ldz g(z, \vec{\sigma}(\vec{\delta r}, \delta t))
\label{eq:dphi} 
\end{equation}
is the wave phase structure function, 
\[
g(z,\vec{\sigma}) = \int_{-\infty}^{+\infty} 
dq_x\int_{-\infty}^{+\infty}dq_y~
\left\{ \right. \hspace{3cm} 
\]
\begin{equation} \left. \hspace{2cm} 
\left[1-\cos(\vec q\cdot\vec{\sigma})\right]P_{n_{\rm
e}}(q_x,q_y,q_z=0; z)\right\},
\label{eq:g} 
\end{equation} 
\begin{equation}
\vec{\sigma}(\vec{r},\delta t) = \frac{z}{L}\vec{\delta r}
+ \vec{V}_{\rm eff} \delta t,
\label{eq:sigma} 
\end{equation}
and 
\begin{equation}
\vec{V}_{\rm eff}(z) = \left(1 - \frac{z}{L}\right) \vec{V}_{\rm p}
+ \frac{z}{L} \vec{V}_{\rm obs}.
\label{eq:veff} 
\end{equation}
Here, $P_{n_{\rm e}}({\bf q};z)$ is the power spectrum of the electron
density fluctuations, ${\bf q}$ is the Fourier-space wave vector, $z$
is the coordinate along the path from the source to the observer, $L$
is the distance between source and observer, $r_{{\rm e}}$ is the
classical radius of the electron, $\lambda$ is the wavelength at which
the observations are taken, $\vec{V}_{\rm p}$ and $\vec{V}_{\rm obs}$
are the velocities of the pulsar and observer, respectively, relative
to the frame of the density fluctuations (which we assume to have
negligible motion) and perpendicular to the line of sight, and $\vec V
_{\rm eff}$ is the effective perpendicular velocity of the sight line
relative to the plasma turbulence.  The power spectrum $P_{n_{\rm e}}$
is taken to depend upon $z$ in what is essentially a two-scale
approximation, since the outer scale of the turbulence, $l$, is taken
to be much smaller than $L$.  The sight line sweeps across a
small-scale density fluctuation in a time short compared to the
Lagrangian correlation time of the density fluctuation, and thus the
intrinsic time evolution of the density fluctuations is ignored.  The
wave phase structure function $D_\phi$ can be thought of as the
mean-square difference between the density-fluctuation-induced phase
increments along one line of sight from the source's position at time
$t- (L/c)$ to the observing location at time $t$ and another line of
sight from the source's position at $t+\delta t - (L/c)$ to the
observing location at time $t+\delta t$.  For strong scattering
[equivalent to $D_\phi  \gg 1$ as $\delta r \rightarrow \infty$], the
intensity correlation function $<I(\vec r,t) I(\vec
r+\vec{\delta r},t+\delta t)>$ is approximately given by (Lotova \&
Chashei 1981, Cordes \& Rickett 1998; see appendix~\ref{ap:deriv} for
a derivation)
\begin{equation} 
<I(\vec r,t) I(\vec r+\vec{\delta r},t+\delta t)>\simeq
<I>^2 \{1 + \exp[-D_\phi(\vec{\delta r},\delta t)\},
\label{eq:gammaf2} 
\end{equation}
where $<I>$ is the mean intensity.

The observed electric field and intensity vary stochastically as the
sight line moves through turbulent density fluctuations in the ISM. To
obtain the visibility and intensity correlation function, the observed
electric field and intensity are averaged over some interval in
time. In the derivation of the formulas quoted above,
it is assumed for simplicity that this averaging
procedure is equivalent to taking an ensemble average over the
electron density fluctuations which determine the observed
electromagnetic fields.  This assumption is justified only if the
integration time is sufficiently long (Goodman \&
Narayan 1989).  The effects of scattering by anisotropic turbulence
for shorter-duration observations are beyond the scope of this paper.

We define orthogonal coordinates $x$ and $y$ in the plane
perpendicular to the line of sight such that in the case of
anisotropic scattering, $y$ is the direction of strongest scattering
(largest angular size).  The length scales of the diffraction pattern
along the $x$ and $y$ directions, given by $l_{{\rm d},x}$ and
$l_{{\rm d},y}$ respectively, and the scintillation time
scale $t_d$ are determined from $D_\phi(\delta x, \delta y, \delta t) \equiv
D_\phi(\vec{\delta r},\delta t)$ by the equations
\begin{eqnarray} 
D_\phi(l_{{\rm d},x}, 0, 0) & = & 1, \label{eq:ldx} \\
D_\phi(0, l_{{\rm d},y}, 0) & = & 1, \label{eq:ldy}  \mbox{ \hspace{0.3cm} and} \\
D_\phi(0,0,t_d) & = & 1. \label{eq:td} 
\end{eqnarray} 
The effective angular size of the source in the $x$ and
$y$ directions is given by 
(Rickett 1990) 
\begin{eqnarray} 
\theta_{s,x} & = & \frac{1}{k l_{{\rm d},x}} \label{eq:thetax}, \mbox{ 
\hspace{0.3cm} and} \\
\theta_{s,y} & = & \frac{1}{k l_{{\rm d},y}} \label{eq:thetay}. 
\end{eqnarray} 

\section{Calculation of the wave phase structure function, diffraction-pattern
length scales, scintillation time scale,
and angular broadening for a GS spectrum of density fluctuations}
\label{sec:calc} 

\begin{figure*}[h]
\vspace{10cm}
\includegraphics{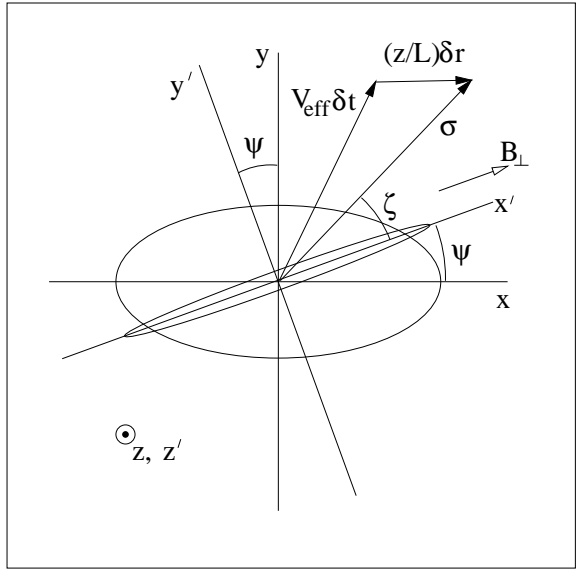}
\caption{\label{fig:xy} The eccentric ellipse along the $x^\prime$
axis corresponds to the diffraction pattern associated
with a thin layer of material with a single magnetic field vector whose
projection in the plane of the sky, $B_\perp$, is along $x^\prime$. The
projection of the density structures would also appear as 
eccentric ellipses elongated along $x^\prime$. The
less eccentric ellipse corresponds to the diffraction pattern resulting from
scattering along the entire line of sight. The $y$ axis is defined
to be along the short axis of the less eccentric ellipse, which is
the long dimension of the corresponding broadened image.}
\end{figure*}

\begin{figure*}[h]
\vspace{11cm}
\includegraphics{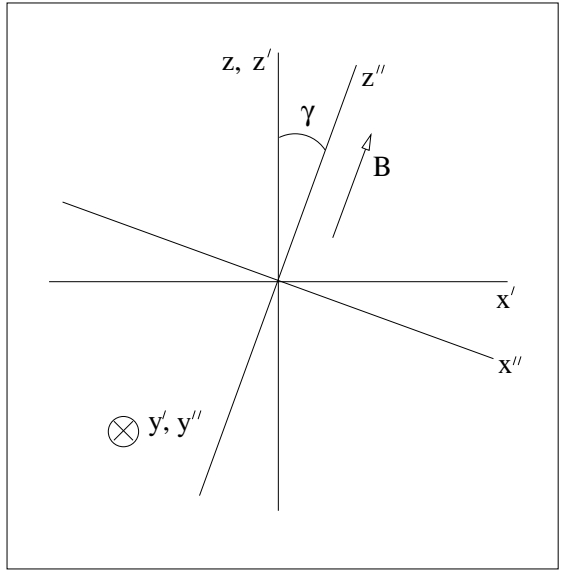}
\caption{The magnetic field is along $z^{\prime\prime}$, which
makes an angle $\gamma$ with the $z$ and $z^\prime$ axes.
\label{fig:xz} }
\end{figure*}

We define orthogonal $x$, $y$, and $z$ axes with $z$ along the line of
sight and $y$ along the direction of maximum angular broadening.  We
define orthogonal $x^\prime$, $y^\prime$, and $z^\prime$ axes with
$z^\prime$ along $z$ and with an angle $\psi(z)$ between $x^\prime$ and
$x$ and between $y^\prime$ and $y$, as in figure~\ref{fig:xy}.  The
value of $\psi$ will vary along the line of sight so that $x^\prime$
is aligned with the projection of $\vec{B}(z)$ in the plane of the
sky. We also define orthogonal $x^{\prime\prime}$, $y^{\prime\prime}$,
and $z^{\prime\prime}$ such that $y^{\prime\prime}$ is along
$y^\prime$ in the original $xy$ plane, and $z^{\prime\prime}$ is
parallel to $\vec{B}(z)$, which is taken to make an angle $\gamma(z)$ with
the $z$ and $z^\prime$ axes, as depicted in figure~\ref{fig:xz}. The
density fluctuations are elongated along $\vec{B}(z)$ and thus
$z^{\prime\prime}$. The density structures and scattering
have the same statistical properties 
for $\gamma = \gamma_0$ and $\gamma = \pi - \gamma_0$.
The separation $\vec{\sigma}$ between
the two lines of sight in the phase-structure-function formula
is taken to make an angle $\zeta(z)$ with respect to the $x^\prime$
axis, as in figure~\ref{fig:xy}.

We assume a GS spectrum for $P_{n_{\rm e}}(\vec{q},z)$ with a sharp
cutoff at scales larger than an outer scale $l$ and at scales smaller
than an inner scale $d$, i.e., if $l^{-1} <q_\perp< d^{-1}$ then
\begin{equation} 
P_{n_{\rm e}}(\vec{q},z) =
 (1/6\pi)\langle \delta n_{\rm e}^2 \rangle l^{-1/3}
q_\perp^{-10/3} f\left(\frac{\displaystyle q_{z^{\prime \prime}}}
{\displaystyle q_\perp^{2/3}l^{-1/3}}
\right), 
\label{eq:p} 
\end{equation}
and if $q_\perp < l^{-1}$ or $q_\perp > d^{-1}$ then
$P_{n_{\rm e}}(\vec{q},z) = 0$, where
\begin{equation}
f(x) \equiv \left\{
\begin{array}{ll}
1 \hspace{0.3cm} & \mbox{ if $|x| < 1$} \\
0 & \mbox{ if $|x| > 1$}
\end{array}
\right.,
\label{eq:f} 
\end{equation} 
and where $q_\perp = (q_{x^{\prime\prime}}^2 + q_{y^{\prime
\prime}}^2)^{1/2}$.  The normalization in equation~(\ref{eq:p}) has
been chosen so that the mean square density fluctuation is given by
$\langle \delta n_{\rm e}^2 \rangle$.

Noting that
\begin{eqnarray} 
q_{x^{\prime\prime}} & = & q_{x^{\prime}}\cos \gamma 
- q_{z^\prime}\sin\gamma, \\
q_{y^{\prime\prime}} & = & q_{y^{\prime}}, \mbox{ \hspace{0.3cm} and}\\
q_{z^{\prime\prime}} & = & q_{x^{\prime}}\sin\gamma
+ q_{z^\prime}\cos\gamma, \\
\end{eqnarray} 
we find from equations~(\ref{eq:p}) and (\ref{eq:f}) 
that $P_{n_{\rm e}}(q_{x^\prime}, q_{y^\prime},
q_{z^\prime}=0, z)$ is nonzero only when
\begin{equation}
|q_{x^\prime}l\sin\gamma|^3 < q_{x^\prime}^2l^2 \cos^2\gamma +q_{y^\prime}^2
l^2.
\label{eq:ineq1} 
\end{equation} 
The fluctuations that dominate radio-wave scattering typically
satisfy $q_{y^\prime} l \gg 1$. When $q_{y^\prime} l \gg 1$
and $\gamma \gg (q_{y^\prime}l)^{-1/3}$ (i.e., the
magnetic field is not too closely aligned with the line
of sight), the first term on the right-hand side of equation~(\ref{eq:ineq1}) 
is negligible compared to the second for all allowable values
of $q_{x^\prime}$, and the 
upper limit on $q_{x^\prime}$ becomes 
\begin{equation}
|q_{x^\prime}l\sin\gamma|^3 \lesssim q_{y^\prime}^2 l^2.
\label{eq:ineq2} 
\end{equation} 
Thus, when $q_{y^\prime} l \gg 1$
and $\gamma \gg (q_{y^\prime}l)^{-1/3}$,
\[
P_{n_{\rm e}}(q_{x^\prime}, q_{y^\prime}, q_{z^\prime}=0, z) \simeq
\hspace{4cm} 
\]
\begin{equation} 
(1/6\pi) \langle \delta n_{\rm e}^2 \rangle l^{-1/3}
q_{y^\prime}^{-10/3} 
f\left(\frac{q_{x^\prime}}{q_{y^\prime}^{2/3} l^{-1/3} 
\csc\gamma}\right).
\label{eq:ap} 
\end{equation}
On the other hand, as $\gamma\rightarrow 0$, $P_{n_{\rm e}}$ becomes
isotropic in $q_{x^\prime}$ and $q_{y^\prime}$.

In order to calculate $D_\phi$ for different cases,
we will first calculate the value of $g(z,\vec{\sigma})$ for various
orientations of the magnetic field direction ${\bf \hat
z^{\prime\prime}}$ relative to the line of sight and the
vector $\vec{\sigma}$ separating the two lines of sight
in the turbulent medium.

\subsection{Value of $g(z,\vec{\sigma}) $ when the magnetic field is along the
line of sight}

When $\sigma$ is in the inertial range ($d \ll
\sigma \ll l$)  and $\gamma \ll (\sigma /l)^{1/3}$,
the dominant contributions to $g(z,\vec{\sigma})$ comes
from values of $q_{x^\prime}$ and $q_{y^\prime}$ satisfying
$q_{x^\prime}^2 + q_{y^\prime}^2 \simeq \sigma^{-2}$, for which
the value of $f$ in $P_{n_{\rm e}}(q_{x^\prime}, q_{y^\prime}, q_{z^\prime}=0, z) $
is equal to 1. Thus, 
when $\sigma$ is in the inertial range ($d \ll
\sigma \ll l$)  and $\gamma \ll (\sigma /l)^{1/3}$,
one can take
\begin{equation}
P_{n_{\rm e}}(q_{x^\prime}, q_{y^\prime}, q_{z^\prime}=0, z) \simeq 
 (1/6\pi)\langle \delta n_{\rm e}^2 \rangle l^{-1/3}
(q_{x^\prime}^2 + q_{y^\prime}^2)^{-5/3},
\end{equation} 
which implies that to lowest order in $(\sigma /l)$
\begin{equation}
g(z,\vec{\sigma})  = \frac{\Gamma(1/3)}{ 2^{10/3} \Gamma(5/3)}
\langle \delta n_{\rm e} ^2 \rangle l^{-1/3} \sigma^{4/3}.
\label{eq:s1} 
\end{equation} 
When $\sigma$ is in the dissipation range ($\sigma \ll d$)
and $\gamma \ll (d/l)^{1/3}$, one finds that to lowest
order in $(\sigma/d)$ and $(d/l)$
\begin{equation}
g(z,\vec{\sigma})  = (1/8) \langle \delta n_{\rm e} ^2\rangle l^{-1/3} d^{-2/3} \sigma^2.
\label{eq:s2} 
\end{equation} 
The values of $g$ in equations~(\ref{eq:s1}) and (\ref{eq:s2}) are
large compared to those that arise from other orientations of
$\vec{B}$ due to the increased coherence length of the elongated
density structures along the line of sight.

\subsection{Value of $g(z,\vec{\sigma}) $ when magnetic field is not along
the line of sight and $\vec{\sigma}$ is not along ${
\bf \hat x^\prime}$}

Let 
\begin{equation} 
u  \equiv  q_{y^\prime} l, 
\end{equation}
\begin{equation} 
a  \equiv \frac{\sigma \cos \zeta}{l} \ll 1, 
\end{equation}
\begin{equation} 
b  \equiv  \frac{\sigma \sin \zeta }{l} \ll 1,
\end{equation} 
and let $u_{\rm min}$ be some value of $u$ 
satisfying
\begin{equation}
1 \ll u_{\rm min} \ll |b|^{-1}.
\label{eq:umin} 
\end{equation} 
For $\gamma \gg u_{\rm min}^{-1/3}$ and for $u>u_{\rm min}$,
$P_{n_{\rm e}}$ is given by equation~(\ref{eq:ap}). Thus,
for $\gamma \gg u_{\rm min}^{-1/3}$, 
the contribution to $g(z,\vec{\sigma}) $ from
$u > u_{\rm min}$ is 
\[
g(z,\vec{\sigma})  = \frac{2 \langle \delta n_{\rm e}^2 \rangle l}{3\pi} 
\int _{u_{\rm min}} ^{l/d} du \,
u^{-10/3} \left[
u^{2/3} \csc\gamma - \hspace{3cm} \right.
\]
\begin{equation}
\left.\hspace{3cm} 
\frac{1}{a} \sin(a u^{2/3}\csc\gamma)
\cos(b u)\right].
\label{eq:g1} 
\end{equation} 
If $d \ll \sigma\sin \zeta\ll l$ 
 and $|a|^{3/2} (\csc \gamma)^{3/2}
\ll |b|$, which implies that
\begin{equation}
|\zeta| \gg (\sigma/l)^{1/2} (\csc\gamma)^{3/2},
\label{eq:cond1} 
\end{equation}
then there exists a $u_1 < l/d$
such that $a u_1^{2/3}\csc\gamma \ll 1$
and $|b u_1| \gg 1$. 
The contribution to $g(z,\vec{\sigma}) $ from $q_{y^\prime} \gg l^{-1}$
is then dominated by values of $u$ in the interval $(u_{\rm min}, u_1)$,
in which the integrand in equation~(\ref{eq:g1}) can be expanded:
\begin{equation}
g(z,\vec{\sigma})  = \frac{2 \langle \delta n_{\rm e}^2 \rangle l}{3\pi}  \int _{u_{\rm min}} ^{u_1} du \,
u^{-8/3} \csc\gamma [1 - \cos(b u)].
\end{equation} 
To lowest order the limits of integration can be changed to
$(0,\infty)$, which implies that the contribution to $g(z,\vec{\sigma}) $ 
from values of $q_{y^\prime}$ much greater than $l^{-1}$ is
\[
g(z,\vec{\sigma})  = \left[\frac{2}{5\Gamma(5/3)}\right]\langle \delta n_{\rm e} ^2\rangle
\sigma^{5/3}
l^{-2/3} \csc\gamma |\sin\zeta|^{5/3}
\hspace{3cm} 
\]
\begin{equation} 
\mbox{ \hspace{3cm} if $d\ll \sigma \sin \zeta \ll l$}.
\label{eq:s3} 
\end{equation} 
Although equation~(\ref{eq:s3}) was derived assuming
$\gamma \gg u_{\rm min}^{-1/3}$, it is
approximately valid under the slightly more general 
condition $\gamma \gg (\sigma |\sin\zeta|/l)^{1/3}$ since
$g(z,\vec{\sigma}) $ is dominated by values of $q_{y^\prime}$ of
order $(\sigma |\sin\zeta|)^{-1}$.

As $\zeta\rightarrow 0$, the contribution to $g(z,\vec{\sigma}) $
from large $q_{y^\prime}$ vanishes, but the contribution
from small $q_{y^\prime}$ does not. Anticipating the results
of the next subsection, we note that
in order for the large-$q_{y^\prime}$
contribution to dominate and for equation~(\ref{eq:s3}) to
be accurate when $\sigma \sin\zeta$ is in
the inertial range, $\zeta$ must satisfy the condition
\begin{equation}
|\zeta| \gg (\sigma/l)^{1/5}(\sin\gamma)^{3/5}
\label{eq:cond2} 
\end{equation} 
as well as equation~(\ref{eq:cond1}).

When $\sigma\sin\zeta$ is in the dissipation range ($\ll d$)
and $\gamma \gg (d/l)^{1/3}$, one finds that the
contribution to $g(z,\vec{\sigma}) $ from $q_{y^\prime}\gg l^{-1}$
is to lowest order
\[
g(z,\vec{\sigma})   = (1/\pi) \langle \delta n_{\rm e}^2 \rangle l^{-2/3}
d^{-1/3} \sigma^2 \sin^2\zeta \csc\gamma
\hspace{1cm} 
\]
\begin{equation} 
\mbox{ \hspace{3cm} if $\sigma \sin\zeta \ll d$}.
\label{eq:s4} 
\end{equation}
In order for the large-$q_{y^\prime}$ contribution
to dominate over the small-$q_{y^\prime}$ contribution,
$\zeta$ must be $\gg (d/l)^{1/6} (\sin\gamma)^{1/2}$.

\subsection{Value of $g(z,\vec{\sigma}) $ when magnetic field is not
along the line of sight but $\vec{\sigma} $ is along~$\bf \hat x^\prime$}
\label{sec:markov} 

When $\sigma \sin\zeta$ is in the inertial range,
$\gamma \gg (\sigma \sin\zeta\,/l)^{1/3}$,
and $|\zeta| \ll (\sigma/l)^{1/5} (\sin\gamma)^{3/5}$,
the dominant contribution to $g(z,\vec{\sigma}) $  in the
Markov approximation comes from $q_{y^\prime}\sim
l^{-1}$, giving in order of magnitude
\begin{equation}
g(z,\vec{\sigma})  \sim  \langle \delta n_{\rm e}^2\rangle \sigma^2 l^{-1}.
\label{eq:s5} 
\end{equation} 
Equation~(\ref{eq:s5}) also holds in the Markov approximation
when $\sigma \sin\zeta$ is in the dissipation range,
$\gamma \gg (d/l)^{1/3}$, and $|\zeta|\ll (d/l)^{1/6} (\sin
\gamma)^{1/2}$. The Markov approximation, however, is formally
invalid for density fluctuations at scales $\sim l$,
where $l$ is on the order of 1 to 100 pc for typical
interstellar conditions. This is because the Markov
approximation assumes that
\begin{equation}
\delta \phi = r_e \lambda l_c \delta n_{\rm e} \ll 1,
\label{eq:markov} 
\end{equation} 
where $\delta \phi$ is the phase increment 
(relative to propagation through the average medium) induced
as the wave propagates through one correlation length $l_c$ of the density
fluctuation $\delta n_{\rm e}$. For the Goldreich-Sridhar spectrum with fractional
density fluctuations of order unity at the outer scale $l$,
one has $\delta n_{\rm e} \sim n_e (l_c/l)^{1/3}$.
Thus, the Markov approximation is valid only for
those density fluctuations with correlation lengths satisfying
\begin{equation}
l_c \ll l^{1/4} \lambda^{-3/4} r_e^{-3/4} n_e^{-3/4}.
\label{eq:markov2} 
\end{equation} 
For $l = 10^{20}$ cm, $\lambda = 30 $ cm, and
$n_e = 0.1 \mbox{ cm}^{-3}$, the Markov approximation is only valid for
$l_c \ll 10^{14}$ cm. Although the Markov approximation
breaks down for the fluctuations at scales larger than 
$\sim 10^{14}$ cm, we will for simplicity 
assume that the Markov-approximation results
are accurate for the large-scale fluctuations in order
of magnitude. In almost all cases, this assumption leads to
the conclusion that the large-scale density fluctuations
are unimportant for scintillation and angular broadening.
The only exception is when all of the density fluctuations are 
aligned in approximately the same direction, as
in section~\ref{sec:thinscreen}. In that case, the large-scale
fluctuations dominate the angular broadening in the direction
parallel to the elongated density structures.

Having determined the values of $g(z,\vec{\sigma}) $ for different 
orientations of the magnetic field relative to $\bf\hat
z$ and $\vec\sigma$, we now turn to  calculations
of $D_\phi$.

\subsection{Value of $D_\phi$ when there is either significant
variation in the magnetic-field direction
$\bf\hat z^{\prime\prime}$ along the line of sight,
or when $\bf\hat z^{\prime\prime}$ is not along the line-of-sight
and $\bf\hat x^\prime$ is not along $\vec\sigma$}
\label{sec:gen}

The $z$ integral appearing in equation~(\ref{eq:dphi}) can be
divided into intervals in which (a) the magnetic field $\vec B$ is
along the line of sight, (b) $\vec B$ is not along the line of sight
and $\bf \hat x^\prime$ is not along $\vec \sigma$, and (c) $\vec B$
is not along the line of sight but $\bf \hat x^\prime$ is along $\vec
\sigma$.  In intervals of type (a), when $\sigma$ is in the inertial
range, the value of the integrand in equation~(\ref{eq:dphi}) is $\sim
(l/\sigma)^{1/3}$ times larger than in intervals of type (b). However,
intervals of type (a) [for which $\gamma < (\sigma/l)^{1/3}$] are
about a factor of $(\sigma/l)^{2/3}$ less common than intervals of
type (b) when the field direction varies
significantly along the line of sight in a random manner, 
and thus intervals of type (a) do not contribute significantly to
$D_\phi$. Intervals of type (c) are much less common than intervals of
type (b), and in addition the integrand in equation~(\ref{eq:dphi}) is
smaller in intervals of type (c) than in intervals of type (b). Thus,
$D_\phi$ is dominated by intervals of type (b), both when
there is significant random variation in the field direction along the
line of sight and when the field direction is fixed in a single
direction of type (b). 

When $d\ll \sigma \ll l$, for most values of $\zeta$ one also has $d
\ll \sigma\sin\zeta \ll l$, the condition under which $g(z,\vec{\sigma}) $ is given
by equation~(\ref{eq:s3}). Thus, when $\sigma $ is in the inertial
range,
\[
D_\phi = \frac{8\pi}{5\Gamma(5/3)} r_e^2 \lambda^2
\int_0^L dz(
\langle \delta n_{\rm e}^2\rangle l^{-2/3} \csc\gamma |\sin\zeta|^{5/3}
\sigma^{5/3})
\]
\begin{equation}  \mbox{ \hspace{3cm} if $d\ll \sigma \ll l$},
\label{eq:d1} 
\end{equation}
where the integration excludes segments of the line of sight 
for which $\gamma \lesssim (\sigma/l)^{1/3}$.
Similar arguments show that when $\sigma$ is in the
dissipation range, the $z$ integral in equation~(\ref{eq:dphi}) 
is dominated by intervals in which $g(z,\vec{\sigma}) $ is given by
equation~(\ref{eq:s4}), giving
\[
D_\phi = 4 r_e^2 \lambda^2 \int_0^L
dz\, (\langle \delta n_{\rm e}^2 \rangle l^{-2/3} d^{-1/3}
\sigma^2 \sin^2\zeta \csc\gamma) 
\]
\begin{equation} 
\mbox{ \hspace{3cm} if $
\sigma \ll d$,}
\label{eq:d2} 
\end{equation} 
where the integration excludes segments of the line of sight 
for which $\gamma \lesssim (d/l)^{1/3}$.

A subtlety not addressed in this paper is that $\sigma$
can be within the inertial range of the turbulence over part of the
line of sight and within the dissipation range over the remainder of
the line of sight. For simplicity it will be assumed that if $\sigma$
is in the inertial range at either the source or at the observer,
then the inertial-range formulas can be used for the entire line
of sight. For example, if $\delta r$ is in the inertial range and
$\delta t =0$, then $\sigma$ is taken to be in the inertial range
along the entire line of sight, although in fact it is in the
dissipation range sufficiently close to the source.

From equations~(\ref{eq:ldx}), (\ref{eq:ldy}), (\ref{eq:td}) and
(\ref{eq:d1}), one finds that for $d \ll \sigma \ll l$,
\begin{equation} 
l_{{\rm d},x}  =  \left( \int_0^1 d\xi L \eta_{\rm gs} \csc\gamma |\sin\psi|^{5/3} \xi^{5/3}
\right)^{-3/5},
\label{eq:gslx} 
\end{equation} 
\begin{equation} 
l_{{\rm d},y}  =  \left( \int_0^1 d\xi L \eta_{\rm gs} \csc\gamma |\cos\psi|^{5/3} \xi^{5/3}
\right)^{-3/5},  \mbox{ \hspace{0.3cm} and}
\label{eq:gsly} 
\end{equation} 
\begin{equation} 
t_d  = \left[ \int_0^1 d\xi L \eta_{\rm gs} \csc\gamma |\sin \zeta|^{5/3}
|(1-\xi) \vec{V}_{\rm p} + \xi \vec{V}_{\rm obs}|^{5/3}|\right]^{-3/5},
\label{eq:gstd} 
\end{equation} 
where $ \eta_{\rm gs} = [8\pi/5\Gamma(5/3)] r_e^2
\lambda^2 \langle \delta n_{\rm e}^2\rangle l^{-2/3}$, $\xi = z/L$, and the
integration excludes segments of the line of sight in which $\gamma
\lesssim (\sigma/l)^{1/3}$.  Equations~(\ref{eq:gslx}) and
(\ref{eq:gsly}) show that the length scale of the diffraction pattern
tends to be dominated by fluctuations near the observer. If $V_{\rm
p}\gg V_{\rm obs}$, then equation~(\ref{eq:gstd}) shows that the time
scale of the diffraction pattern tends to be dominated by fluctuations
near the source.

\subsection{Special case: homogeneous extended
medium with a statistically isotropic random magnetic field direction}

Let us suppose that the magnetic-field direction varies
randomly in an isotropic manner
along the line of sight, and that $\langle \delta n_{\rm e}^2\rangle$,
$l$, and $d$ are constant along the line of sight. We have already
assumed that $l\ll L$, which implies that in the $z$ integral
in equation~(\ref{eq:d1}) we can average over the direction of
$\vec B$ while holding $z$ constant to a good approximation.
With
\begin{eqnarray} 
\alpha & \equiv & \frac{32\pi^2}{5\Gamma(5/3)} \int_0^{\pi/2}
d\zeta (\sin\zeta)^{5/3}, \\
\xi & \equiv & \frac{z}{L}, \mbox{ \hspace{0.3cm} and} \\
H & \equiv & (r_e^2 \lambda^2 \langle \delta n_{\rm e}^2\rangle L)^{-1},
\end{eqnarray} 
one finds that
\[
D_\phi = \alpha H^{-1} l^{-2/3} \int_0^1 d\xi\,
|(\vec{\delta r} - \vec V _{\rm p} \delta t + \vec V_{\rm obs}
\delta t)\xi + \vec V_{\rm p} \delta t|^{5/3}
\]
\begin{equation} 
\mbox{ \hspace{3cm} if $d\ll \sigma \ll l$.}
\end{equation}
When $\delta t = 0$, or $V_{\rm p} = V_{\rm obs} = 0$, 
\begin{equation}
D_\phi = (3/8) \alpha H^{-1} l^{-2/3} \delta r^{5/3}.
\mbox{ \hspace{0.3cm} if $d\ll \delta r \ll l$.}
\end{equation} 
The diffraction pattern in this case is isotropic, with
\begin{equation}
l_{{\rm d},x} = l_{{\rm d},y} = \left(\frac{8}{3\alpha}\right)^{3/5} H^{3/5} l^{2/5}
\mbox{ if $d\ll l_{{\rm d},x} = l_{{\rm d},y} \ll l$.}
\end{equation}

When $\sigma $ is in the dissipation range, one finds
that
\[
D_\phi = 4\pi^2 H^{-1} l^{-2/3} d^{-1/3}\hspace{0.3cm}  \times
\]
\begin{equation} 
[(1/3)|\vec{\delta r} + \vec V_{\rm obs}\delta t -\vec V_{\rm p} \delta t|^2
+\vec{\delta r} \cdot \vec{V}_{\rm p}\delta t + \vec{V}_{\rm p}\cdot
\vec{V}_{\rm obs}\delta t^2] \mbox{ \hspace{0.3cm} if $\sigma \ll d$},
\end{equation}
\begin{equation}
l_{{\rm d},x} = l_{{\rm d},y} =  \left(\frac{3}{4\pi^2}\right)^{1/2} H^{1/2} l^{1/3} d^{1/6}
\mbox{ \hspace{0.3cm} if $l_{{\rm d},x} = l_{{\rm d},y} \ll d$},
\end{equation}
and
\[
t_d =  \left(\frac{3}{4\pi^2}\right)^{1/2}
\frac{H^{1/2} l^{1/3} d^{1/6}}{(V_{\rm obs}^2 + 
\vec V_{\rm obs}\cdot \vec{V}_{\rm p} + V_{\rm p}^2)^{1/2}}
\]
\begin{equation} 
\mbox{ \hspace{3cm} if $l_{{\rm d},x} = l_{{\rm d},y} \ll d$}.
\end{equation}

\subsection{Special case:
one thin scattering screen with a single direction of $\vec{B}$}
\label{sec:thinscreen} 

In this section, it is assumed that all of the electron
density fluctuations lie within a screen of
thickness $\Delta z\ll l \ll L$. The magnetic field direction
thus does not rotate very much within the screen, and the
density fluctuations are aligned in the same direction.
The screen is taken to be a distance $z$ from the source.
Upon defining 
\begin{equation}
h \equiv (r_e^2 \lambda ^2 \Delta z
\langle \delta n_{\rm e}^2 \rangle)^{-1}
\end{equation}
(for $\lambda = 10$ cm, $\Delta z = 10^{18}$ cm,  and
$\langle \delta n_{\rm e}^2 \rangle = 1 \mbox{ cm}^{-6}$,
$ h = 1.3 \times 10^5$ cm), one finds that 
\begin{equation}
l_{{\rm d},y} = \left(\frac{5\Gamma(5/3)}{8\pi}\right)^{3/5} h^{3/5}
l^{2/5}(\sin\gamma)^{3/5} \frac{L}{z}
\mbox{ \hspace{0.3cm} if $d\ll z l_{{\rm d},y}/L \ll l$}.
\end{equation}
To obtain an approximate value for
$l_{{\rm d},x}$, one must use equation~(\ref{eq:s5}) to obtain
\begin{equation}
l_{{\rm d},x} \sim h^{1/2} l^{1/2} \frac{L}{z}.
\label{eq:lx2} 
\end{equation} 
Equation~(\ref{eq:lx2}) holds whether $z l_{{\rm d},x}/L$ is in the
dissipation range or the inertial range.  As discussed in section
\ref{sec:markov}, we are unable to determine how $l_{{\rm
d},x}$ changes with $\gamma$  as $\csc\gamma$ becomes
large. When $\csc\gamma$ is of order unity (e.g., $\lesssim 4$)
the axial ratio
of the anisotropic diffraction pattern and broadened image
is given by
\begin{equation}
\frac{\theta_{s,y}}{\theta_{s,x}}=
\frac{l_{{\rm d},x}}{l_{{\rm d},y}} \sim \left(\frac{l}{h}\right)^{1/10} 
\mbox{ \hspace{0.3cm} if $d \ll zl_{{\rm d},y}/L \ll l$},
\end{equation} 
ignoring factors of order unity.
The axial ratio of the density structures in the Goldreich-Sridhar
spectrum that dominate the
diffraction pattern is 
\begin{equation}
[l/(zl_{{\rm d},y}/L)]^{1/3} \sim \left(\frac{l}{h}\right)^{1/5}
\mbox{ \hspace{0.3cm} if $d \ll zl_{{\rm d},y}/L \ll l$}.
\end{equation}
Thus, for $\csc\gamma$ of order unity, the axial ratio of the
diffraction pattern is roughly the square root of the axial ratio of
the turbulent structures that dominate the scattering at the scale of
the diffraction pattern (i.e. turbulent structures with
$q_{y^\prime}^{-1} \sim zl_{{\rm d},y}/L$). It should be noted,
however, that this result follows from applying the Markov
approximation to the outer-scale density fluctuations, for which the
approximation is formally invalid.  There exist other anisotropic
power spectra for which the axial ratio of the diffraction pattern
arising from a thin screen equals the axial ratio of the density
structures. For example, if $P_{n_{\rm e}}(q_{x^{\prime \prime}},
q_{y^{\prime \prime}}, q_{z^{\prime \prime}}) \propto [q_{x^{\prime
\prime}}^2 + q_{y^{\prime \prime}}^2 + (R^{\prime \prime})^2
q_{z^{\prime \prime}}^2]^{-\beta/2}$ where $R^{\prime\prime}$ and
$\beta$ are constants, $R^{\prime\prime} \gg 1$, and $\beta < 4$, and
if $\csc\gamma$ is of order 1, then the axial ratio of the diffraction
pattern is $\sim R^{\prime\prime}$ (Backer \& Chandran 2000).
Similarly, Narayan \& Hubbard (1988) showed that if the
two-dimensional power spectrum of the phase fluctuations in a thin
scattering screen is $\propto [(R^{\prime\prime})^2 q_x^2 +
q_y^2]^{-\beta/2}$ with $\beta < 4$, then the axial ratio of the
associated diffraction pattern is $R^{\prime\prime}$.  The key
difference between these axial ratios and the axial ratios for a GS
spectrum is the following: although the image broadening along the $y$
direction for a GS spectrum is dominated by highly anisotropic
small-scale fluctuations, the image broadening along the $x$ direction
for a GS spectrum is dominated by the large-scale fluctuations.  Since
the image broadening along $x$ is enhanced relative to the amount of image
broadening arising solely from the small-scale density structures, 
the image is less anisotropic than the density structures.

The scintillation time scale $t_d$ is given by
\[
t_d = \left[\frac{5\Gamma(5/3)}{8\pi}\right]^{3/5}
\frac{h^{3/5} l^{2/5} |\csc\zeta|(\sin\gamma)^{3/5}}
{V_{\rm eff}} 
\]
\begin{equation} 
\mbox{ \hspace{3cm} if $d\ll zV_{\rm eff} t_d/L \ll l$}.
\end{equation} 
However, as $\zeta \rightarrow 0$, $t_d$ does not approach $\infty$
but instead reaches a maximum value when $D_\phi$ is determined
using equation~(\ref{eq:s5}):
\begin{equation}
t_{d,\rm max} \sim \frac{h^{1/2}l^{1/2}}{V_{\rm eff}}.
\label{eq:tdmax} 
\end{equation}
If $|\vec{V}_{\rm eff}$ is constant but $\zeta$ varies continuously
through a $360^\circ$ interval 
during
the course of either the earth's orbit around the sun or the orbit
of a binary pulsar, then the ratio of the maximum to minimum
scintillation times when $\csc\gamma$ is of order unity is given by
\begin{equation}
\frac{t_{d,\rm max}}{t_{d,\rm min}} \simeq \left(\frac{l}{h}\right)^{1/10}.
\end{equation} 

For $zl_{{\rm d},y}/L$ in the dissipation range, one finds that
\begin{equation}
l_{{\rm d},y} = (1/2) h^{1/2} l^{1/3} d^{1/6} (\sin\gamma)^{1/2}\frac{L}{z}
\mbox{ \hspace{0.3cm} if $zl_{{\rm d},y}/L \ll d$}.
\end{equation} 
Since equation~(\ref{eq:lx2}) applies in the dissipation
range, the axial ratio of the diffraction pattern 
and broadened image for $\csc\gamma$ of order  unity is given by
\begin{equation}
\frac{\theta_{s,y}}{\theta_{s,x}}=
\frac{l_{{\rm d},x}}{l_{{\rm d},y}} \sim \left( \frac{l}{d}\right)^{1/6}
\mbox{ if $z l_{{\rm d},y}/L \ll d$},
\end{equation}
which again,  is approximately
the square root of the axial ratio of the turbulent structures
at scale $d$ that dominate the scattering for $z l_{{\rm d},y}/L \ll d$.
The scintillation time scale is
\begin{equation}
t_d = \frac{h^{1/2} l^{1/3} d^{1/6} (\sin\gamma)^{1/2}\csc\zeta}
{2 V_{\rm eff}} \mbox{ \hspace{0.3cm} if $z l_{{\rm d},y}/L \ll d$.}
\end{equation}
As $\zeta\rightarrow 0$, $t_d$ stops increasing after reaching the
value given in equation~(\ref{eq:tdmax}).  Again, if $V_{\rm eff}$ is
constant but $\zeta$ varies continuously through a $360^\circ$
interval during the course of either the earth's orbit around the sun
or the orbit of a binary pulsar, then the ratio between the maximum and
minimum scintillation times when $\csc\gamma$ is of order unity is
given by 
\begin{equation}
\frac{t_{d,\rm max}}{t_{d,\rm min}} \simeq \left(\frac{l}{d}\right)^{1/6}.
\end{equation}

\subsection{Special case: thin scattering screen in which the
direction of magnetic field in plane of sky
rotates through an angle $\Delta \psi$}
\label{sec:rot} 

\begin{figure*}[h]
\vspace{10cm}
\includegraphics{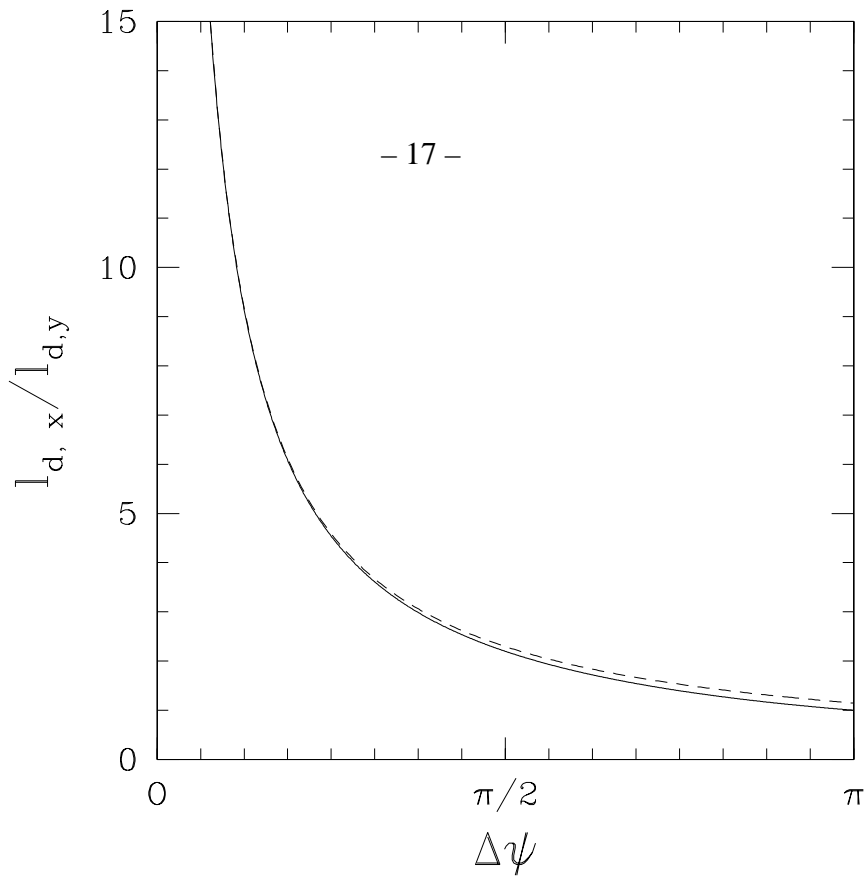}
\caption{Axial ratio of the diffraction pattern and 
broadened image of a point source when the magnetic field
direction in the plane of the sky rotates through an angle
$\Delta \psi$ within the scattering screen. The solid line
represents equation~(\ref{eq:ani1}), and the dashed line
represents equation~(\ref{eq:ani2}), which is derived assuming
$|\Delta \psi | \ll 1$.
\label{fig:figure1}  }
\end{figure*}

In this section, it is assumed that the scattering medium
extends from $z=z_1$ to $z=z_1 + \Delta z$,
 with $\Delta z \ll L$. It is assumed that
$\langle \delta n_{\rm e}^2 \rangle$, $l$, and $\gamma$ 
are constant within the screen, that $\sigma$ is in
the inertial range of the density fluctuations, and 
that $\gamma \gg (\sigma\sin \zeta/l)^{1/3}$.
The angle $\psi$ between the
projection of the magnetic field in the plane
of the sky and the $x$ axis is given by
\begin{equation}
\psi = - \frac{\Delta \psi}{2} + \frac{(z-z_1)\Delta \psi }{\Delta z}.
\label{eq:psi} 
\end{equation}
The value of $\Delta \psi$ is conceptually equivalent to $\Delta z/l$.
It is assumed that
\begin{equation}
\Delta \psi \gg R^{-1/2},
\label{eq:psimin} 
\end{equation}
where $R$ is the axial ratio of the small-scale
density structures that dominate the scattering.
This means that for any direction of $\vec{\sigma}$,
portions of the line of sight of type (c) in 
section~\ref{sec:gen}  (for which $\vec{\sigma}$ is
nearly along the direction of ${\bf B}$ in the plane
of the sky) contribute a negligible amount to $D_\phi$,
and $D_\phi$ is given by equation~(\ref{eq:d1}). 
From this it follows that 
\begin{equation}
\frac{l_{{\rm d},x}}{l_{{\rm d},y}} =
\left[\frac{\displaystyle \int_0^{\Delta \psi/2} d\psi\,|\cos \psi|^{5/3}}
{\displaystyle \int_0^{\Delta \psi/2} d\psi\,|\sin \psi|^{5/3}}\right]^{3/5}.
\label{eq:ani1} 
\end{equation}
When $|\Delta \psi | \ll 1$, equation~(\ref{eq:ani1}) reduces to
\begin{equation} 
\frac{l_{{\rm d},x}}{l_{{\rm d},y}} \simeq
\frac{2(8/3)^{3/5}}{\Delta \psi} \simeq \frac{3.6}{\Delta \psi}.
\label{eq:ani2} 
\end{equation}
Equations~(\ref{eq:ani1}) and (\ref{eq:ani2}) are plotted in
figure~\ref{fig:figure1}. If $|\vec{V}_{\rm eff}|$ is constant during
one orbit of the earth around the Sun or one orbit of a binary
pulsar, then the ratio of the maximum to minimum scintillation times
$t_{\rm d, max}/t_{\rm d, min}$ during the course of one
orbit is also given by 
\begin{equation}
\frac{t_{\rm d, max}}{t_{\rm d, min}} =
\left[\frac{\displaystyle  \int_0^{\Delta \psi/2} d\psi\,|\cos \psi|^{5/3}}
{\displaystyle  \int_0^{\Delta \psi/2} d\psi\,|\sin \psi|^{5/3}}\right]^{3/5},
\label{eq:ani3} 
\end{equation}
and when $|\Delta \psi | \ll 1$, 
\begin{equation} 
\frac{t_{\rm d, max}}{t_{\rm d, min}} \simeq
\frac{2(8/3)^{3/5}}{\Delta \psi} \simeq \frac{3.6}{\Delta \psi}.
\label{eq:ani4} 
\end{equation}
The scintillation time reaches its maximum value when
the sight line moves parallel to $\hat{x}$, the direction
in which the density fluctuations are
most elongated, and reaches its minimum when the sight line moves along
$\hat{y}$.

\section{Summary of results}
\label{sec:conc}

In this paper, we assume a GS spectrum (Goldreich \& Sridhar 1995)
of electron density fluctuations in the ISM.
We then calculate general formulas for the wave phase
structure function,  visibility, angular broadening,
diffraction-pattern length scales, and scintillation time scale for arbitrary
distributions of turbulence along the line of sight, and specialize
these formulas to idealized cases.
Our main results are as follows:

(1) Unless the magnetic field is closely aligned with the line of sight
over a much greater fraction of the line of sight than is expected for
random magnetic fields, the scaling of the wave phase structure
function $D_\phi$ with baseline $\delta r$ is $D_\phi \propto (\delta
r)^{5/3}$ for $\delta r$ in the inertial range of the turbulence, and $D_\phi
\propto(\delta r)^2$ for $\delta r$ smaller than the dissipation scale
or inner scale of the turbulence, just as for an isotropic Kolmogorov
spectrum of density fluctuations.

(2) If the magnetic field were closely aligned with the line of sight
[within an angle $\sim (\delta r/l)^{1/3}$ if $\delta r$ is in the
inertial range, and $\sim (d/l)^{1/3} $ if $\delta r$ is in the
dissipation range] over more than a minimum-threshold fraction of the
line of sight, scattering would be dominated by that fraction of the
line of sight and $D_\phi $ would be proportional to $(\delta
r)^{4/3}$ if $\delta r$ were in the inertial range, and to $(\delta
r)^2$ if $\delta r$ were smaller than the dissipation scale. The
minimum-threshold fraction is small [$\sim (\delta r/l)^{1/3}$ if
$\delta r $ is in the inertial range, or $\sim(d/l)^{1/3}$ if $\delta r$
is in the dissipation range], but much larger than expected for
randomly varying magnetic fields. The increase in scattering when
$\vec{B}$ becomes aligned along the line of sight is due to the
increased coherence length of the elongated density structures along
the line of sight.

(3) Angular broadening is most anisotropic when scattering occurs
within a thin screen in which the magnetic field has a single
direction. In this case, the axial ratio of the broadened image of a
point source in the Markov approximation is approximately the square
root of the axial ratio of the density fluctuations that dominate the
scattering, provided that $\csc \gamma $ is of order unity
(e.g. $\lesssim 4$), where $\gamma$ is the angle between the magnetic
field and the line of sight.  We have not found the dependence of the
axial ratio on $\gamma$ when $\csc\gamma$ is large, although we do
show that if $\gamma = 0$ the broadened image is isotropic.

(4) If the scattering medium is confined to a homogeneous thin screen
of thickness $\Delta z \ll L$, and if 
the direction of the magnetic field in the plane of the sky
rotates through an angle $\Delta \psi \gg R^{-1/2}$ along the line of sight
from one side of the screen to the other,  
where $R$ is the axial ratio of the density fluctuations that dominate
the scattering, then the axial ratio of the observed image of a point
source is $\simeq 3.6/\Delta \psi$. [This formula is derived assuming
$\Delta \psi \ll 1$, but closely approximates (to within
15\%) the more general formula, equation~(\ref{eq:ani1}), 
for $\Delta \psi$ as large as $\pi$.]  This
indicates that even a moderate variation in the field direction within
the scattering medium can dramatically reduce the axial ratio of an
angularly broadened imaged.  If the field direction along the line of
sight to a source varies significantly, the resulting
angular-broadening anisotropy 
is determined by the amount of variation in the field direction
rather than by the degree of anisotropy of the density fluctuations.

(5) When the magnetic field in the scattering medium has a single
direction or rotates through a relatively small angle $\Delta \psi$,
the scintillation time is longest when the sight line moves along the
direction of greatest elongation of the electron density fluctuations
in the scattering medium, and shortest when the sight line moves
orthogonal to the direction of greatest elongation of the density
structures. This gives rise to an observable modulation of the
scintillation time during the course of either the Earth's orbit
around the Sun, or of a binary pulsar's orbit. When the speed of the
sight line in the plane of the sky $V_{\rm eff}$ in the scattering
medium (assumed to be a thin screen) is constant during the orbit, the
ratio of the maximum to minimum scintillation times $t_{\rm d,
max}/t_{\rm d, min}$ is the same as the axial ratio of the broadened
image of a point source, when the magnetic field has a single direction
in the thin screen as well as when the magnetic field rotates through
an angle $\Delta \psi$. If the field direction along the line of sight
to a source varies significantly, the degree of variation in the
scintillation time is determined by the amount of variation in the
field direction rather than by the degree of anisotropy of the density
fluctuations.

(6) The diffraction-pattern length scales and angular broadening
are more sensitive to fluctuations near the earth than to fluctuations
near the pulsar. The time scale for diffractive
scintillations is more sensitive to fluctuations near the pulsar
if the pulsar speed significantly exceeds the speed of the
observer with respect to the interstellar turbulence.

\acknowledgements{We thank Steve Spangler, Bob Mutel,
Sridhar Seshadri, Jason Maron, and Barney Rickett for helpful discussions. 
This research has been supported by NSF grant AST-9820662 at UC Berkeley,
and by NSF grant AST-0098086 and DOE grants DE-FG02-01ER54658 and
DE-FC02-01ER54651 at the University of Iowa.}

\appendix

\section{Derivation of equations for visibility and intensity correlation function
for a moving observer and moving point source}
\label{ap:deriv} 

In this appendix we derive equations~(\ref{eq:vis}) through (\ref{eq:gammaf2})
for the visibility and intensity correlation function for a moving
point source and moving observer. In several places, the derivation is
identical to the corresponding calculation for plane waves incident
upon a turbulent medium and observer that are at rest, and the reader
will be referred to Lee \& Jokipii (1975a,b) for the details.

We work in the rest frame of the source, and 
start with the scalar wave equation for electromagnetic waves
propagating in a cold non-magnetized plasma (Faraday rotation
is ignored), 
\begin{equation}
\nabla^2 E-
\frac{1}{c^2} \frac{\partial ^2 E}{\partial t^2}
- \frac{\omega_p^2}{c^2} E = 0,
\label{eq:disp1.5} 
\end{equation} 
where 
\begin{equation}
\omega_p^2 = \frac{ 4\pi n_e e^2}{m_e}.
\label{eq:op} 
\end{equation} 
We look for solutions that are approximately spherical waves,
\begin{equation}
E = u e^{i(kr - \omega t)},
\label{eq:eu} 
\end{equation}
with
\begin{equation}
u \rightarrow 1 \mbox{ \hspace{0.3cm} as \hspace{0.3cm} } r\rightarrow 0,
\label{eq:bc} 
\end{equation} 
where $(r,\theta,\phi)$ are spherical coordinates centered on the source, and,
in the notation of Lee \& Jokipii (1975a),
\begin{equation} 
k^2  =  \frac{\omega ^2 \epsilon_0}{c^2}, \hspace{0.5cm} 
\epsilon_0  =  \langle \epsilon_\ast \rangle, \mbox{ \hspace{0.3cm} and
\hspace{0.3cm} }
\epsilon_\ast  =  1 - \frac{\omega_p ^2}{\omega^2}. 
\end{equation} 
The angled brackets denote an ensemble average over the turbulent
density fluctuations.
We also define 
\begin{equation}
\epsilon = -\frac{4\pi \delta n_{\rm e} r_e}{k^2}
\end{equation} 
to be the fractional fluctuations in the dielectric constant $\epsilon_\ast$,
where $r_e$ is the classical electron radius $e^2/(m_e c^2)$, 
so that 
\begin{equation}
\epsilon_\ast = \epsilon_0 (1 + \epsilon).
\end{equation} 
We make the ``parabolic approximations,''
\begin{equation} 
k\frac{\partial u }{\partial r} \gg  \frac{\partial ^2 u}{\partial r^2}
\hspace{0.5cm} \mbox{ and} \hspace{0.5cm} \omega  \frac{\partial u }{\partial t} \gg  
\frac{\partial ^2 u}{\partial t^2},
\label{eq:par} 
\end{equation} 
and using equations~(\ref{eq:disp1.5}) through (\ref{eq:par}) 
find that
\begin{equation}
2ik\left( \frac{\partial u}{\partial r}  + \frac{1}{v_g} \frac{\partial u}{
\partial t}\right) + \nabla_\perp^2 u + \epsilon k^2 u = 0,
\label{eq:disp2} 
\end{equation} 
where
\begin{equation}
v_g = c\sqrt{\epsilon_0},
\end{equation} 
and
\begin{equation}
\nabla_\perp^2 = \frac{1}{r^2}\left[ \frac{1}{\sin \theta}\frac{\partial }
{\partial \theta}\left(\sin \theta \frac{\partial }{\partial \theta}\right)
+ \frac{1}{\sin^2\theta}\frac{\partial^2}{\partial\phi^2}\right].
\label{eq:np} 
\end{equation} 
We then transform coordinates from $(r,\theta,\phi,t)$ to
$(z, \theta, \phi, s)$, where
\begin{equation}
z = r \hspace{0.3cm} \mbox{ and} \hspace{0.3cm} s = r - v_g t.
\end{equation} 
Points with a constant $s$ are all those values of $r$ and $t$ associated
with the wavefront that arrives at the observer at the time $t= (L-s)/v_g$,
where $L$ is the distance from source to observer.
In terms of $z$ and $s$ and the shorthand notation 
defined in table \ref{tab:sh}, equation~(\ref{eq:disp2}) can be written
\begin{equation}
2ik\frac{\partial u_1}{\partial z} + \nabla_{\perp,1}^2 u_1+ \epsilon_1 k^2 u_1 = 0.
\label{eq:disp3} 
\end{equation} 
Equation~(\ref{eq:disp3})  is of the same form as equation (7) of
Lee \& Jokipii (1975a).
\begin{table}
\begin{center}
\begin{tabular}{cc}
\hline
Abbreviation & Meaning \\
\hline \hline
\\
$u_i$ & $u(z, \theta_i, \phi_i, s_i)$ \\
$u_i^\prime$ & $u(z^\prime, \theta_i, \phi_i, s_i)$ \\
$\epsilon_i$ & $\epsilon(z, \theta_i, \phi_i, s_i)$ \\
$\epsilon_i^\prime$ & $\epsilon(z^\prime, \theta_i, \phi_i, s_i)$ \\
$\nabla_{\perp,i}^2$ & $\displaystyle \frac{1}{z^2} \left[
\frac{1}{\sin\theta_i} \frac{\partial}{\partial \theta_i}\left(
\sin\theta_i \frac{\partial}{\partial \theta_i}\right)
 + \frac{1}{\sin^2\theta_i}\frac{\partial^2}{\partial \phi_i^2}\right]$ \\
\\
\hline
\end{tabular}
\caption{Shorthand notation, where $i=1, 2, 3,$ or 4. \label{tab:sh}}
\end{center}
\end{table}
Taking the complex conjugate
of equation~(\ref{eq:disp3}), and evaluating the terms at
$\theta_2$, $\phi_2$, and $s_2$, one can write
\begin{equation}
2ik \frac{\partial u_2^\ast}{\partial z} - \nabla_{\perp, \,2}^2
u_2^\ast - \epsilon_2 k^2 u_2^\ast = 0.
\label{eq:disp4} 
\end{equation} 
Multiplying equation~(\ref{eq:disp3}) by $u_2^\ast$ and
equation~(\ref{eq:disp4}) by $u_1$ and adding, taking the
ensemble average, and noting that for statistically
homogeneous density fluctuations $\nabla_{\perp,1}^2
\langle u_1 u_2^\ast\rangle
 = \nabla_{\perp,2}^2
\langle u_1 u_2^\ast\rangle$, we find that
\begin{equation}
2ik \frac{\partial }{\partial z} \langle u_1 u_2^\ast\rangle
+ k^2 \langle (\epsilon_1 - \epsilon_2)u_1 u_2^\ast\rangle = 0.
\label{eq:adisp} 
\end{equation} 
We now make the Markov approximation, in which we 
assume that the correlation length of the density fluctuations $l_\epsilon$
is much smaller than the length $l_u$ along the line of sight
over which $u$ changes significantly.
We then follow a procedure analogous to the one
described in appendix A of
Lee \& Jokipii (1975a) to show that
\begin{equation}
 \langle (\epsilon_1 - \epsilon_2)u_1 u_2^\ast\rangle =
\frac{ik \langle u_1 u_2^\ast\rangle}{2} \int_{-\infty}^z
dz_1 (\langle \epsilon_1 \epsilon_1^\prime \rangle +
\langle \epsilon_2 \epsilon_2^\prime \rangle -
\langle \epsilon_1 \epsilon_2^\prime \rangle -
\langle \epsilon_2 \epsilon_1^\prime \rangle).
\label{eq:mark} 
\end{equation} 

To evaluate the right-hand side of
equation~(\ref{eq:mark}),  we first
define $\alpha$ according to the equation
\begin{equation}
\theta = \frac{\pi}{2} + \alpha.
\end{equation}
We restrict our attention to values of $\alpha$ and $\phi$ 
satisfying
\begin{equation}
|\alpha| \ll 1 \mbox{ \hspace{0.3cm} and \hspace{0.3cm} } |\phi| \ll 1.
\end{equation} 
We then introduce 
Cartesian coordinates 
$(x, y)$ in the plane perpendicular to the
line of sight, with
\begin{equation}
x = z\alpha \mbox{ \hspace{0.3cm} and \hspace{0.3cm}} y = z\phi.
\label{eq:xy} 
\end{equation} 
(Note: the $xyz$ coordinates used here are not related to $\theta$
and $\phi$ in the conventional way. For example, the $z$ 
axis corresponds not to $\theta=0$, but instead to $\theta = \pi/2$
and $\phi = 0$.)
We also make a two scale approximation along the line of sight,
dividing the dependence of the variables on the radial coordinate
into two parts:
a dependence on $Z$ reflecting variations over scales $\sim l_\epsilon$,
and a dependence on $z$ reflecting variations over scales $\gg l_\epsilon$.
The integral in equation~(\ref{eq:mark}) effectively extends over a distance
$\sim l_\epsilon \ll z$, and so in calculating $x$ and $y$ for $\epsilon_1^\prime$
and $\epsilon_2^\prime$ we can use $z$ in place of $z^\prime$ in 
equation~(\ref{eq:xy}) with only small error.
Defining $\vec{x} \equiv x\hat {x}  + y\hat{y}  +Z\hat{z} $, and assuming
statistical homogeneity and stationarity, we can write
\begin{equation}
\langle \delta n_{\rm e}(\vec{x}_1,t_1;z) \delta n_{\rm e}(\vec{x}_2, t_2;z) \rangle
\equiv Q(\vec{x}_1 - \vec{x}_2, t_1 - t_2; z).
\label{eq:Q} 
\end{equation}
We assume that the density fluctuations are static in the rest frame
of the turbulent medium, which is 
reasonable since the Lagrangian correlation time of
the density fluctuations that dominate scintillation 
is small in the GS theory compared to the time for the line
of sight to sweep across such a fluctuation. This gives
\begin{equation}
Q(\vec{x}, t; z) = G(\vec{x} - \vec{U}_m t; z),
\label{eq:G} 
\end{equation} 
where $\vec{U}_m$ is the uniform velocity of the turbulent medium with
respect to the source, and $G(\vec{x}; z)$ is the spatial correlation function
of the density structures in the rest frame of the
turbulent medium, which is the inverse Fourier transform of the
density power spectrum~$P_{n_{\rm e}}$:
\begin{equation}
G(\vec{x}; z) = \int d^3 q \, e^{i\vec{q}\cdot \vec{x}}
P_{n_{\rm e}}(\vec{q}; z).
\label{eq:RP} 
\end{equation} 
Defining
\begin{equation}
H_{ij} \equiv \int_{-\infty}^z dz^\prime \langle \epsilon_i \epsilon_j^\prime
+ \epsilon_j \epsilon_i^\prime\rangle,
\label{eq:H0} 
\end{equation} 
we find that
\begin{equation}
H_{ij} = 
\frac{32 \pi^3 r_e^2}{k^4} \int d^3 q \,P_{n_{\rm e}}(\vec{q}; z)
e^{i\vec{q}\cdot \vec{p}_{ij}}
\delta\left( \vec{q} \cdot
\left[\hat{z} - \frac{\vec{U}_m}{v_g}\right] \right) ,
\label{eq:H1} 
\end{equation}
where
\begin{equation}
\vec{p}_{ij} = z(\alpha_i - \alpha_j) \hat{x} + z(\phi_i - \phi_j) \hat{y}
+ \frac{(s_i - s_j)\vec{U}_m}{v_g} .
\label{eq:pij} 
\end{equation} 
The presence of the $\vec{U_m}/v_g$ term in the delta function in
equation~(\ref{eq:H1}) is due to aberration---in the frame of
reference of the turbulent fluctuations, the wave appears to be moving
in a direction slightly offset from the line of sight. Since $U_m \ll
v_g $, this effect is small and can be ignored, and the delta
function can be written $\delta (k_z)$.  Equation~(\ref{eq:H1}) thus
reduces to
\begin{equation}
H_{ij} = 
\frac{32 \pi^3 r_e^2}{k^4} \int \int dq_x dq_y \,P_{n_{\rm e}}(q_x, q_y, q_z=0; z)
e^{i\vec{k}\cdot \vec{\sigma}_{ij}},
\label{eq:H2} 
\end{equation}
where
\begin{equation}
\vec{\sigma}_{ij} = z(\alpha_i - \alpha_j) \hat{x} + z(\phi_i - \phi_j) \hat{y}
+ \frac{(s_i - s_j)\vec{U}_{m,\perp}}{v_g} ,
\label{eq:sigmaij} 
\end{equation} 
and where $\vec{U}_{m\,\perp}$ is the component of $\vec{U}_m$ perpendicular to the
line of sight.

Equations~(\ref{eq:adisp}), (\ref{eq:mark}),
(\ref{eq:H0}),  and~(\ref{eq:H2}) imply that
\begin{equation}
\frac{\partial }{\partial z} \langle u_1 u_2^\ast\rangle
= - 2 \pi r_e^2 \lambda^2 g(z, \vec{\sigma}_{12}) \langle u_1 u_2^\ast\rangle.
\label{eq:diffeq} 
\end{equation} 
If the correlated observations are taken at locations that are
separated in the observer's rest frame 
by a displacement $\vec{\delta r} = \delta x \hat{x}
+ \delta y \hat{y} + \delta z \hat{z}$  and at times that
are separated by a time interval $\delta t$, then
\begin{eqnarray} 
L(\alpha_1 - \alpha_2) & = &\delta x + U_{\rm obs,x}\delta t,
\label{eq:alph} 
\\
L(\phi_1 - \phi_2) & = & \delta y + U_{\rm obs,y} \delta t, \mbox{ \hspace{0.3cm} and}\\
s_1 - s_2 & = & -v_g(t_1 - t_2)  .
\label{eq:s9} \\
\end{eqnarray} 
Because $u$ changes much more slowly along the line of sight
than across the line of sight, $\delta z$ and $U_{\rm obs, z}$ can be neglected. Defining
 $V_p$ and $V_{\rm obs}$ as the velocities of the source  and observer 
perpendicular to the line of sight measured
in the rest frame of the turbulent medium, one has 
\begin{equation}
\vec{U}_{m,\perp} = - \vec{V}_p ,
\end{equation}
\begin{equation} 
\vec{U}_{\rm obs,\perp} = - \vec{V}_p + \vec{V}_{\rm obs},
\label{eq:veltr} 
\end{equation} 
and
\begin{equation}
\vec{\sigma}_{12} = \frac{z}{L} (\vec{\delta r} + \vec{V}_{\rm obs} \delta t)
+ \left( 1 - \frac{z}{L}\right) \vec{V}_p \delta t.
\label{eq:sig2} 
\end{equation} 
Integrating equation~(\ref{eq:diffeq}) and using equations~(\ref{eq:bc}) 
and~(\ref{eq:sig2}),
one obtains equation~(\ref{eq:vis}).

To derive the intensity correlation function, we define
\begin{equation}
\Gamma_{2,2} = \langle u_1 u_2 u_3^\ast u_4^\ast\rangle .
\end{equation} 
Using the same procedure used to derive equation~(\ref{eq:adisp}),
one obtains
\begin{equation}
2ik\frac{\partial\Gamma_{2,2} }{\partial z} 
+ (\nabla_{\perp,1}^2 + \nabla_{\perp,2}^2 
- \nabla_{\perp,3}^2 - \nabla_{\perp,4}^2) \Gamma_{2,2}
+ k^2 \langle (\epsilon_1 + \epsilon_2 - \epsilon_3 - \epsilon_4)
u_1 u_2 u_3^\ast u_4^\ast\rangle = 0.
\label{eq:i1} 
\end{equation} 
Using the Markov approximation, we find that
\begin{equation}
\langle (\epsilon_1 + \epsilon_2 - \epsilon_3 - \epsilon_4)
u_1 u_2 u_3^\ast u_4^\ast\rangle =
\frac{ik\Gamma_{2,2}}{2}
(2H_{11} + H_{12} + H_{34} - H_{13} - H_{14} - H_{23} - H_{24} ),
\label{eq:mark2} 
\end{equation} 
which gives 
\begin{equation}
\frac{\partial \Gamma_{2,2} } {\partial z} = 
\frac{i}{2k}\left(
\nabla_{\perp,1}^2 + \nabla_{\perp,2}^2 
- \nabla_{\perp,3}^2 - \nabla_{\perp,4}^2\right)  \Gamma_{2,2}
+ \frac{k^2 \Gamma_{2,2}}{4}\left(
2H_{11} + H_{12} + H_{34} - H_{13} - H_{14} - H_{23} - H_{24} \right).
\label{eq:int2} 
\end{equation}
To an excellent approximation, 
\begin{equation}
\nabla_{\perp,i}^2 \simeq \frac{1}{z^2}\left(
 \frac{\partial^2}{\partial \alpha_i^2} +
\frac{\partial^2}{\partial \phi_i^2}\right),
\end{equation}
and thus equation~(\ref{eq:int2}) is equivalent to equation (8)
of Lee \& Jokipii (1975b), if the effects of aberration are ignored.
Lee \& Jokipii's (1975b) derivation of the approximate
intensity correlation function in the limit of strong scattering
can therefore be applied to equation~(\ref{eq:int2}) to obtain
equation~(\ref{eq:gammaf2}).

\section{References}

Armstrong, J. W., Coles, W. A., Kojima, M., \& Rickett, B. J. 1990,
ApJ, 358, 685

Armstrong, J. W., Rickett, B. J., \& Spangler, S. R. 1995, ApJ, 443, 209

Backer, D. C., \&  Chandran, B. 2000, ApJ, submitted

Bhattacharjee, A., \& Ng, C. S. 2000, ApJ, submitted

Bondi, M., Padrielli, L., Gregorini, L., Mantovani, F., Shapirovskaya, N.,
\& Spangler, S. R. 1994, Astron. \& Astrophys., 287, 390

Cho, J., \& Vishniac, E. 2000, ApJ, 539, 274

Coles, W., Frehlich, R., Rickett, B, \& Codona, J. 1987, ApJ, 315, 666

Frail, D., Diamond, P, Cordes, J., \& van Langevelde, H. 1994, ApJ 427, L43

Ghosh, S., \& Goldstein, M. 1997, J. Plasma Phys., 57, 129

Goldreich, P., \& Sridhar, S. 1995, ApJ, 438, 763 

Goldreich, P., \& Sridhar, S. 1997, ApJ, 485, 680

Goodman, J., \& Narayan, R. 1989, MNRAS, 238, 995

Higdon, J. C. 1984, ApJ, 285, 109

Higdon, J. C. 1986, ApJ, 309, 342

Iroshnikov, P. 1963, A Zh, 40, 742

Kraichnan, R. H. 1965, Phys. Fluids, 8, 1385

Lee, L. C., \& Jokipii, J. R. 1975a, ApJ, 196, 695

Lee, L. C., \& Jokipii, J. R. 1975b, ApJ, 202, 439

Lithwick, Y., \& Goldreich, P. 2001, in press

Lotova, N., and Chashei, I. 1981, Sov. Astron., 25, 309

Maron, J., \& Goldreich, P 2001, ApJ, in press

Matthaeus, W., Oughton, S., \& Ghosh, S., Phys. Rev. Lett., 81, 2056

Molnar, L. A., Mutel, R. L., Reid, M. J., \& Johnston, K. J. 1995, ApJ, 438, 708

Montgomery, D., \& Matthaeus, W. 1995, ApJ, 447, 706

Montgomery, D., \& Turner, L. 1981, Phys. Fluids, 24, 825

Mutel, R. L., \&  Lestrade, J.-F. 1990, 349, L47

Narayan, R., Anantharamaiah, K. R., \& Cornwell, T. J. 1990, MNRAS, 241, 403

Narayan, R., \& Hubbard, W. B. 1988, ApJ, 325, 503

Oughton, S., Priest, E., \& Matthaeus, W. 1994, J. Fluid Mech., 280, 95

Rickett, B. J. 1990, Ann. Rev. Astron. Astrophys., 28, 561

Shebalin, J. V., Matthaeus, W., \& Montgomery, D. 1983, J. Plasma Phys., 29, 525

Spangler, S. R. 1999, ApJ, 522, 879

Spangler, S. R., \& Cordes, J. 1998, ApJ, 505, 766

Sridhar, S., \& Goldreich, P. 1994, ApJ, 432, 612

Trotter, A. S., Moran, J. M., Rodriguez, L. F. 1998, ApJ, 493, 666

van Langevelde, H., Frail, D., Cordes, J., \& Diamond, P. 1992, ApJ, 396, 686

Wilkinson, P, Narayan, R., \& Spencer, R. 1994, MNRAS, 269, 67

\end{document}